\documentclass{article}
\usepackage{iclr2026_conference,times}
\usepackage{graphicx}
\usepackage{amsmath,amsfonts}
\usepackage{booktabs}
\usepackage{hyperref}
\usepackage{enumitem}
\usepackage{tikz}
\usepackage{pgfplots}
\pgfplotsset{compat=1.18}
\usepgfplotslibrary{fillbetween}
\usetikzlibrary{decorations.pathreplacing} % for braces
\usepackage{hyperref}

\usepackage{amssymb}   % additional symbols
\usepackage{amsfonts} % math fonts
\usepackage{mathtools} % extension of amsmath (recommended)
\usepackage{xcolor}
\definecolor{bluesky}{RGB}{220,235,250}

\usepackage{array}      % for column formatting
\usepackage{colortbl}   % for \rowcolor
\usepackage{xcolor}     % for color definitions (gray!, blue!)

\definecolor{jasper}{rgb}{0.84, 0.23, 0.24}
\definecolor{jazzberry}{rgb}{0.65, 0.04, 0.37}
\definecolor{deepskyblue}{rgb}{0.0, 0.75, 1.0}
\newcommand{\bluesky}{\textcolor{deepskyblue}{Blue Sky}}

\title{Evidence of a Cognitive Shift in AI Education: How Students Are Rethinking Human Intelligence?}
\author{%
  Islem Rekik \\ 
  BASIRA Lab, Imperial-X (I-X) and Department of Computing\\
  Imperial College London, London, United Kingdom\\
  \texttt{\{i.rekik\}@imperial.ac.uk}, \url{https://basira-lab.com/}
}

\iclrfinalcopy % Uncomment for camera-ready version, but NOT for submission.
\begin{document}
\maketitle

\begin{abstract}
Perceptions of intelligence influence how learners evaluate and rely on artificial intelligence (AI) systems. Despite rapid advances in AI capabilities, little is known about how sustained exposure to AI tools affects students’ valuation of human (or natural) intelligence (HI) relative to artificial (or machine) intelligence. This \bluesky{} paper reports a longitudinal classroom response to a poll comparing the perceived importance of AI versus HI at the opening lecture of AI-focused courses between 2020 and 2026, spanning undergraduate and MSc programs in computer science. Responses from 471 students across technical (Machine Learning, Deep Graph Learning) and design-oriented (Design Thinking for AI) courses were analyzed. We identify four recurring phases: \textbf{(1) Hype, (2) Distrust, (3) Trust, and (4) Dependency}. Early measurements in 2020 slightly favored AI over HI. From 2024 onward, preferences consistently shifted toward human intelligence across all MSc cohorts, reaching 65\% (a 12 percentage-point increase from 2025) in a technical AI course and 90\% (a 36 percentage-point increase from 2025) in a design-oriented AI course by 2026. These observations suggest a gradual reappraisal of human intelligence as AI becomes a routine tool that may affect learners’ autonomy and epistemic agency. We conclude this perspective paper by offering introspective insights into a cognitive shift from favoring artificial intelligence toward prioritizing natural intelligence.
\end{abstract}

\section{Introduction}
Artificial (or machine) intelligence (AI) has achieved unprecedented performance across perception, language, and reasoning learning tasks and benchmarks \citep{xiao2025fast,yue2025does,chow2025physbench,shim2025tooldial}. In particular, recent advances in large language models (LLMs) have demonstrated strong capabilities in program synthesis and structured problem solving, enabling models to generate executable code, reason over intermediate steps, and iteratively refine solutions. Empirical studies show that these models can solve a wide range of programming tasks by decomposing problems into modular subcomponents and leveraging learned representations of programming patterns \citep{chen2021codex, li2022competition, wei2022chain}. While these results highlight the effectiveness of implicit, sequence-based reasoning, they have also motivated interest in alternative architectures that aim to endow models with more explicit and persistent cognitive structures. In this context, a new breed of graph neural networks (GNNs) has recently emerged, claiming to equip neural systems with higher-level cognitive capacities such as visual and auditory memory through a dual formalization of graph and reservoir computing \citep{soussia2025multi, soussia2025coggnn}. 

These developments suggest a complementary paradigm for structured reasoning and memory, increasing the appeal of such models in \emph{educational settings}, particularly in computer science, where students use them for coursework, code generation, and conceptual clarification. Empirical studies document that students increasingly integrate large language models such as ChatGPT into their academic workflows for writing, programming, explanation, and problem solving across technical disciplines \citep{bernabei2023students, kasneci2023chatgpt}. As students move from abstract expectations about AI to sustained, everyday interaction with generative models, their lived experience may reshape how they interpret and value the notion of “intelligence.” Therefore, \emph{we hypothesize that increased reliance on AI systems can shift students’ learning priorities toward acquiring proficiency in AI tools, potentially at the expense of investing time and effort in developing their human intelligence (HI)}, including metacognitive skills--that is, the ability to monitor and regulate one’s own cognitive processes \citep{fleur2021metacognition}.

This paper is motivated by an unexpected empirical observation drawn from longitudinal classroom studies in computer science conducted between 2020 and 2026. After several years in which students displayed mixed impressions about the importance of AI versus HI, a marked shift emerged in 2026: \textbf{across two distinct AI-focused courses, students independently reasserted the primacy of human intelligence.}

This \bluesky{} paper therefore explores the following prompting question: \emph{What changes in student cognition, self-perception, or lived experience with AI precipitated this reversal?} We emphasize that our analysis is exploratory in nature, offering preliminary introspections and interpretive insights rather than quantitative evidence.

%This raises a broader question relevant to human-centered AI research: can educational settings, where people actively learn alongside AI systems, reveal early changes in how intelligence is understood and valued?

%%% ================
\section{Cognitive shifts: A Longitudinal Classroom Signal Interpretation}
\subsection{Prompt and protocol}

Since 2020, at the start of multiple AI-focused courses in computer science (e.g., \href{https://www.youtube.com/watch?v=HyWmnlahXAA&list=PLug43ldmRSo1LDlvQOPzgoJ6wKnfmzimQ}{Machine Learning}, \href{https://github.com/basiralab/DGL}{Deep Graph Learning}, and Design Thinking for AI), MSc and UG (undergraduate) students answered anonymously the following question:

\begin{center}
\colorbox{bluesky!35}{%
\parbox{0.9\linewidth}{%
\centering
\emph{We should invest more effort in improving:}\\[0.3em]
\emph{machine intelligence \quad or \quad human intelligence}
}}
\end{center}

\begin{figure}[t]
\centering
\begin{tikzpicture}
\begin{axis}[
    width=0.95\linewidth,
    height=6.2cm,
    ymin=0, ymax=100,
    xmin=2019.5, xmax=2026.5,
    xtick={2020,2024,2025,2026},
    xticklabel style={/pgf/number format/1000 sep=},
    ylabel={Average student vote (\%)},
    xlabel={Year},
    ymajorgrids=true,
    grid style=dashed,
    tick label style={font=\small},
    label style={font=\small},
    legend style={at={(0.5,-0.25)},anchor=north,legend columns=2},
    clip=false
]

% -------------------------------------------------------
% TOP ANNOTATIONS (braces + bold labels), like the example
% -------------------------------------------------------
% Note: Using axis coordinates so placement is stable.
% Braces sit slightly above the top axis line; labels are bold and centered.

% -------------------------------------------------------
% TOP ANNOTATIONS (fixed overlap between III and IV)
% -------------------------------------------------------

% I. Hype (around 2020)
\draw[decorate, decoration={brace, amplitude=4pt}, thick, color=jasper]
  (axis cs:2019.75,103) -- (axis cs:2020.25,103);
\node[anchor=south, font=\bfseries\small, text=jazzberry]
  at (axis cs:2020.0,106) {I. Hype};

% II. Distrust (around 2024)
\draw[decorate, decoration={brace, amplitude=4pt}, thick, color=jasper]
  (axis cs:2023.65,103) -- (axis cs:2024.35,103);
\node[anchor=south, font=\bfseries\small, text=jazzberry]
  at (axis cs:2024.0,106) {II. Distrust};

% III. Trust (shift label slightly LEFT)
\draw[decorate, decoration={brace, amplitude=4pt}, thick, color=jasper]
  (axis cs:2024.85,103) -- (axis cs:2025.15,103);
\node[anchor=south, font=\bfseries\small, text=jazzberry]
  at (axis cs:2024.92,106) {III. Trust};

% IV. Dependency (shift label slightly RIGHT)
\draw[decorate, decoration={brace, amplitude=4pt}, thick, color=jasper]
  (axis cs:2025.85,103) -- (axis cs:2026.15,103);
\node[anchor=south, font=\bfseries\small, text=jazzberry]
  at (axis cs:2026.08,106) {IV. Dependency};

% -------------------------------------------------------
% (1) RAW MARKERS: HI and AI per course offering (sparse)
%     NOTE: Multiple points can occur within the same year.
% -------------------------------------------------------

% HI markers (shiny purple)
\addplot[
    only marks,
    color=violet,
    mark=o,
    thick
]
coordinates {
    (2020,41)
    (2020,53)
    (2024,58)
    (2025,53)
    (2025,54)
    (2026,65)
    (2026,90)
};

% AI markers (light red)
\addplot[
    only marks,
    color=red!60,
    mark=square,
    thick
]
coordinates {
    (2020,59)
    (2020,47)
    (2024,42)
    (2025,47)
    (2025,46)
    (2026,35)
    (2026,10)
};

% -------------------------------------------------------
% Labels: centered, but non-overlapping
% Strategy:
%   - HI labels go ABOVE marks (anchor=south)
%   - AI labels go BELOW marks (anchor=north)
%   - Within the same year, stagger with yshifts
% -------------------------------------------------------

% ---- 2020 ----
% HI (above)
\node[anchor=south, yshift=3pt]
  at (axis cs:2020,41) {\fontsize{5}{6}\selectfont ML (MSc, 2020)};
\node[anchor=south, yshift=10pt]
  at (axis cs:2020,53) {\fontsize{5}{6}\selectfont LfD (UG, 2020)};
% AI (below)
\node[anchor=north, yshift=-3pt]
  at (axis cs:2020,59) {\fontsize{5}{6}\selectfont ML (MSc, 2020)};
\node[anchor=north, yshift=-10pt]
  at (axis cs:2020,47) {\fontsize{5}{6}\selectfont LfD (UG, 2020)};

% ---- 2024 ----
% HI (above)
\node[anchor=south, yshift=3pt]
  at (axis cs:2024,58) {\fontsize{5}{6}\selectfont DGL (MSc, 2024)};
% AI (below)
\node[anchor=north, yshift=-3pt]
  at (axis cs:2024,42) {\fontsize{5}{6}\selectfont DGL (MSc, 2024)};

% ---- 2025 ----
% HI (above) - two close points, stagger more
\node[anchor=south, yshift=3pt]
  at (axis cs:2025,53) {\fontsize{5}{6}\selectfont DGL (MSc, 2025)};
\node[anchor=south, yshift=10pt]
  at (axis cs:2025,54) {\fontsize{5}{6}\selectfont DT4AI (MSc, 2025)};
% AI (below) - two close points, stagger more
\node[anchor=north, yshift=-3pt]
  at (axis cs:2025,47) {\fontsize{5}{6}\selectfont DGL (MSc, 2025)};
\node[anchor=north, yshift=-10pt]
  at (axis cs:2025,46) {\fontsize{5}{6}\selectfont DT4AI (MSc, 2025)};

% ---- 2026 ----
% HI: keep the top label INSIDE the plot to avoid clipping
\node[anchor=south, yshift=3pt]
  at (axis cs:2026,65) {\fontsize{5}{6}\selectfont DGL (MSc, 2026)};
\node[anchor=north, yshift=-3pt] % below the 90% point to avoid hitting the top border
  at (axis cs:2026,90) {\fontsize{5}{6}\selectfont DT4AI (MSc, 2026)};
% AI: stagger; keep the 10% label above so it doesn't collide with x-axis area
\node[anchor=north, yshift=-3pt]
  at (axis cs:2026,35) {\fontsize{5}{6}\selectfont DGL (MSc, 2026)};
\node[anchor=south, yshift=3pt] % above the 10% point for readability
  at (axis cs:2026,10) {\fontsize{5}{6}\selectfont DT4AI (MSc, 2026)};

% -------------------------------------------------------
% (2) YEARLY AVERAGES: shaded band = min..max, line = mean
%     Means are UNWEIGHTED across offerings within each year.
% -------------------------------------------------------

% HI band (min..max)
\addplot[name path=HIupper, draw=none] coordinates {
    (2020,53) (2024,58) (2025,54) (2026,90)
};
\addplot[name path=HIlower, draw=none] coordinates {
    (2020,41) (2024,58) (2025,53) (2026,65)
};
\addplot[violet!20] fill between[of=HIlower and HIupper];

% HI mean line
\addplot[color=violet, thick] coordinates {
    (2020,47.0)
    (2024,58.0)
    (2025,53.5)
    (2026,77.5)
};

% AI band (min..max)
\addplot[name path=AIupper, draw=none] coordinates {
    (2020,59) (2024,42) (2025,47) (2026,35)
};
\addplot[name path=AIlower, draw=none] coordinates {
    (2020,47) (2024,42) (2025,46) (2026,10)
};
\addplot[red!20] fill between[of=AIlower and AIupper];

% AI mean line
\addplot[color=red!60, thick] coordinates {
    (2020,53.0)
    (2024,42.0)
    (2025,46.5)
    (2026,22.5)
};

\legend{Human intelligence, Artificial intelligence}
\end{axis}
\end{tikzpicture}

\caption{
 Longitudinal trends in student preferences for investing in human intelligence (HI) versus artificial intelligence (AI) across AI-focused courses from 2020 to 2026. These are based on the student responses to the prompting poll \textcolor{deepskyblue}{\emph{``We should invest more effort in improving: human intelligence or artificial intelligence.''}}. From 2024 onward, the yearly mean preference for HI exceeds that of AI, with the largest within-year divergence observed in 2026 due to two course offerings exhibiting sharply contrasting valuations. We interpret this temporal pattern as suggestive of a directional cognitive and self-perceptual shift, in which sustained, hands-on experience with AI systems is accompanied by a re-evaluation of the role and value of human intelligence. Top braces annotate four hypothesized phases (\textbf{Hype, Distrust, Trust, Dependency}) to support interpretation of such temporal dynamics. 
}
\label{fig:hi_ai_trend_mean_band}
\end{figure}
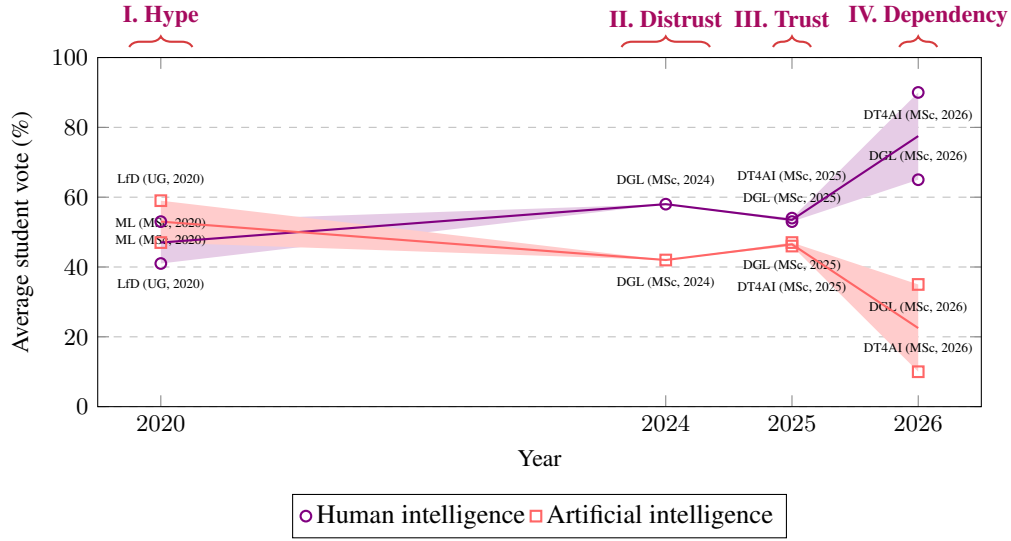

Responses were collected in the opening lecture, before technical instruction, to measure \emph{pre-instructional} beliefs. Table~\ref{tab:hi_ai_preferences} summarizes students’ responses to this question across AI-focused courses between 2020 and 2026, reporting the relative preference for investing in HI versus AI. \textbf{Fig}~\ref{fig:hi_ai_trend_mean_band} illustrates the cumulative trend in student preferences for investing in HI versus AI. Each point corresponds to a distinct course offering and is annotated by year and acronym. 

In 2020, preferences are mixed, with AI favored in a technically oriented MSc course and HI slightly favored in an undergraduate data course. By 2024, HI is consistently preferred in an advanced technical MSc course (Deep Graph Learning). In 2025, preferences remain near parity but continue to slightly favor HI across both technical and design-oriented AI courses. In 2026, a pronounced shift toward HI emerges, most notably in a Design Thinking for AI course, where 90\% of students favor investment in human intelligence. 

Although the observations are sparse and not population-representative, they reveal a clear directional shift in expressed preferences over time. We interpret this pattern as suggestive of a broader cognitive and self-perceptual reorientation, in which students gradually re-evaluate the role and value of human intelligence as their direct, sustained experience with AI systems increases.

\begin{table}[t]
\centering
\caption{
Student preferences for investing in human intelligence (HI) versus artificial intelligence (AI),
collected at the opening lecture of AI-focused courses between 2020 and 2026 in computer science. ITU: Istanbul Technical University. ICL: Imperial College London.
}
\label{tab:hi_ai_preferences}
\begin{tabular}{lllccc}
\toprule
Year & Course & Acronym (institution) & \#Students & HI (\%) & AI (\%) \\
\midrule
2020 & \href{https://www.youtube.com/watch?v=HyWmnlahXAA&list=PLug43ldmRSo1LDlvQOPzgoJ6wKnfmzimQ}{Machine Learning} (MSc)        & ML (ITU)        & 51  & 41 & 59 \\
2020 & \href{https://www.youtube.com/watch?v=HyWmnlahXAA&list=PLug43ldmRSo1LDlvQOPzgoJ6wKnfmzimQ}{Learning from Data} (UG)       & LfD (ITU)      & 89  & 53 & 47 \\
2024 & \href{https://www.youtube.com/watch?v=gQRV_jUyaDw&list=PLug43ldmRSo14Y_vt7S6vanPGh-JpHR7T}{Deep Graph Learning} (MSc)     & DGL (ICL)     & 106 & 58 & 42 \\
2025 & \href{https://www.youtube.com/watch?v=gQRV_jUyaDw&list=PLug43ldmRSo14Y_vt7S6vanPGh-JpHR7T}{Deep Graph Learning} (MSc)     & DGL  (ICL)     & 113 & 53 & 47 \\
2025 & Design Thinking for AI (MSc)  & DT4AI (ICL)     & 26  & 54 & 46 \\
2026 & \href{https://www.youtube.com/watch?v=gQRV_jUyaDw&list=PLug43ldmRSo14Y_vt7S6vanPGh-JpHR7T}{Deep Graph Learning} (MSc)     & DGL  (ICL)     & 65  & 65 & 35 \\
2026 & Design Thinking for AI (MSc)  & DT4AI (ICL)    & 21  & 90 & 10 \\
\bottomrule
\end{tabular}
\end{table}

% =========================================================
\section{Interpreting the Four Phases of Student Perceptions of Intelligence}
% =========================================================

\textbf{Fig}~\ref{fig:hi_ai_trend_mean_band} reveals a clear directional change in how students value human versus artificial intelligence over time. Because this pattern is derived from a single forced-choice question administered to classroom cohorts, it should not be interpreted as a population-level estimate. Instead, we treat it as an \emph{interpretive signal} that motivates reflection on how sustained interaction with generative AI may influence students’ self-perceptions, learning priorities, and epistemic values. The four phases annotated in the figure are therefore not proposed as universal or sequential psychological stages, but as conceptual lenses for organizing these shifts, which may overlap or reoccur as students’ experience with AI deepens. We detail the four hypothesized stages below.

\textbf{\textcolor{jazzberry}{I–Hype Phase: ``The New Hype'' ($HI < AI$)}}  
In the initial phase, students are highly motivated to explore novel machine learning models and devote substantial intellectual effort to engaging with the emerging hype surrounding artificial intelligence. During this period, AI models are often perceived as intelligent machines, leading students to temporarily deprioritize investment in human intelligence and to downplay its importance (Fig.~\ref{fig:hi_ai_trend_mean_band}). This shift does not reflect an explicit rejection of human intelligence, but rather a reduced awareness of its central role in the learning process.

\textbf{\textcolor{jazzberry}{II–Distrust Phase: ``The Reality Check'' ($HI > AI$)}}  As students accumulate direct experience with AI systems, particularly LLMs and generative AI, initial enthusiasm is often tempered by encounters with failure, including hallucinated content, shallow reasoning, and misalignment with evaluative criteria. Such experiences commonly prompt a corrective shift toward skepticism. Verification becomes more salient but also more cognitively demanding, as students must actively monitor and validate AI outputs rather than accept them at face value. In this phase, renewed investment in human intelligence (HI) is often reactive, driven less by intrinsic valuation of human cognition than by concerns about academic risk, reliability, and accountability.

\textbf{\textcolor{jazzberry}{III–Trust Phase: ``The Intellectual Co-Pilot'' ($HI \sim AI$)}}  With continued advances in AI tools and their reasoning and problem-solving capabilities, many students develop a more calibrated understanding of AI as an assistive rather than substitutive technology. In this phase, AI is integrated into learning workflows as a cognitive scaffold, supporting brainstorming, clarification, and exploratory reasoning, while responsibility for evaluation and understanding remains firmly with the learner. Crucially, this orientation foregrounds metacognitive regulation: students actively decide when and how to use AI, assess the quality of its outputs, and integrate them into their own reasoning. This phase reflects a pragmatic equilibrium in which artificial and human intelligence are valued for their complementary roles.

\textbf{\textcolor{jazzberry}{VI- Dependency Phase: ``The Skill Atrophy Risk'' ($HI >> AI$)}} The final phase reflects a more ambiguous dynamic. As AI systems become deeply embedded in everyday academic practice, some students report difficulty initiating or structuring reasoning without external prompts. Rather than indicating a loss of capability, this pattern suggests a redistribution of cognitive effort, in which generation and organization are partially externalized to the system. Over time, however, such reliance may alter how students perceive their own cognitive competence, leading HI to be valued less for productivity and more for its protective function—maintaining standards of evidence, coherence, and epistemic ownership.

Taken together, these phases provide a lens for interpreting the longitudinal shift observed in \textbf{Fig.}~\ref{fig:hi_ai_trend_mean_band}. The increasing preference for human intelligence can be understood as a cognitive and self-perceptual recalibration shaped by sustained, lived experience with AI systems. The following section offers an introspective discussion of the implications of this shift.

% =========================================================
\section{Conclusion and Outlook}
% =========================================================

This \bluesky{} paper reports a longitudinal pedagogical signal observed across six years of AI-focused education: a gradual but directional shift in how students value artificial intelligence relative to human intelligence. Early cohorts consistently favored investment in AI, \textbf{yet by 2025--2026 this pattern reversed}. In 2026, across two independent cohorts and distinct course contexts, a clear majority of students ranked HI as more important than AI at the beginning of their AI courses. Rather than interpreting this reversal as an abrupt backlash, we argue that it reflects a cumulative cognitive and self-perceptual re-evaluation shaped by sustained, everyday interaction with generative AI systems.

We interpret this trend through four recurring phases of student perception (\textbf{\textcolor{jazzberry}{Hype, Distrust, Trust, and Dependency}}), which together provide a conceptual lens for understanding how exposure to AI reshapes learning priorities over time. Initial enthusiasm frames AI as a shortcut that temporarily obscures the learner’s own cognitive role \textcolor{jazzberry}{($HI < AI$)}. Subsequent encounters with system failures prompt skepticism and a reactive reinvestment in human intelligence \textcolor{jazzberry}{($HI > AI$)}. With continued use, many students arrive at a calibrated equilibrium in which AI functions as a cognitive scaffold rather than a substitute \textcolor{jazzberry}{($HI \sim AI$)}. Deeper integration, however, introduces the risk of dependency, where human intelligence is revalued less for efficiency and more for its protective function—judgment, verification, and epistemic ownership \textcolor{jazzberry}{($HI \gg AI$)}. 

The longitudinal shift observed in \textbf{Fig.}~\ref{fig:hi_ai_trend_mean_band} is consistent with increasing time spent in the latter phases, particularly Trust and early Dependency, suggesting that students may be learning to separate capability from responsibility. This, in turn, suggests a growing recognition that while AI can generate, predict, and optimize, human intelligence remains the locus of accountability and epistemic agency. Another explanation for the post-2023 shift is that students are becoming aware that AI outputs can \emph{shape} human judgement, not merely assist it—an expression of epistemic self-defense\footnote{Epistemic self-defense refers to the set of cognitive, metacognitive, and social practices individuals use to protect their beliefs, knowledge, and reasoning processes from deception, error, manipulation, or epistemic harm. It is the ability to defend one’s epistemic agency in environments where information is abundant, persuasive, or unreliable.}.

We emphasize the limits of our evidence. The signal reported here is derived from a single forced-choice question administered to opportunistic classroom cohorts and does not support population-level inference or causal claims. In particular, the prompt \textcolor{deepskyblue}{\emph{``We should invest more effort in improving: machine intelligence or human intelligence?''}} necessarily simplifies a complex, multidimensional relationship into a binary choice, potentially amplifying contrast and obscuring more nuanced or complementary views. Cohort composition, institutional context, and contemporaneous social or technological factors may also contribute to the observed pattern. Nevertheless, the stability of early trends followed by a consistent directional shift motivates systematic investigation. At minimum, these observations suggest that educational settings may serve as early indicators of broader cognitive and cultural changes in \emph{how intelligence is understood and valued}.

\textbf{Why this matters.} If the observed reversal reflects a genuine cognitive reappraisal, it points toward several research directions at the intersection of human-centered AI, education, and human–AI co-learning. These include examining how users’ beliefs about AI competence and intent shape reliance and oversight; designing interfaces that scaffold reasoning without displacing epistemic agency; developing hybrid systems that expose structure, rationale, and uncertainty to support human meaning-making; and rethinking curricula to explicitly preserve and assess core human capabilities such as judgment, synthesis, and moral reasoning under pervasive AI assistance.

The observed cognitive shifts are also consistent with broader empirical evidence on when and how human--AI combinations are effective. A large-scale meta-analysis by \citet{vaccaro2024combinations} shows that hybrid human--AI systems outperform either humans or AI alone \emph{only under specific conditions}--most notably when human judgment, oversight, and contextual reasoning remain actively engaged. \textbf{When human agency is diminished or displaced, performance gains from human--AI synergy often disappear or reverse.}

\colorbox{bluesky!35}{%
\parbox{1\linewidth}{%
\textbf{Insight.} From this \bluesky{ }perspective, the goal of education is not to resist AI, but to guide learners toward the Trust Phase while actively preventing a slide into Dependency. Success lies in enabling students to use AI to reach higher levels of abstraction, meta-cognition and creativity without relinquishing the foundational cognitive skills that constitute natural intelligence. Understanding how and when this balance is achieved remains an open and urgent question for future research.
}}

% =========================================================
\section*{Broader Impact Statement}
% =========================================================
Understanding how learners’ valuations of human and artificial intelligence evolve over time can inform the design of human-aligned AI systems, educational policy, and interaction paradigms that preserve epistemic agency, mitigate overreliance, and support meaningful human oversight in AI-mediated environments. This also challenges prevailing definitions of intelligence and how its various forms are expressed.

% =========================================================
\section*{LLM Usage Disclosure}
% =========================================================
In accordance with ICLR policy, we disclose that ChatGPT was used solely for drafting assistance and language refinement. All research questions, interpretations, data collection, and conceptual contributions and formalizations were developed by the authors, who remain fully responsible for the content of this submission.

\bibliography{ref}
\bibliographystyle{iclr2026_conference}
\end{document}